\documentclass{aa}   
\usepackage{graphicx}
\usepackage{txfonts}
\usepackage{natbib}
\usepackage{natbib}

\def\bi {\begin{itemize}}
\def\ei {\end{itemize}}
\def\intgr{\textit {INTEGRAL}}

\def\xmm{\textit {XMM-Newton}}
\def\thc {3C~273}
\def\deg {$^\circ$}
\usepackage{color}
\definecolor{red}{rgb}{0.7,0,0}
\definecolor{blue}{rgb}{0,0,0.7}

\begin{document}
   \title{2003--2005 \textit{INTEGRAL} and {\it XMM-Newton} observations of 3C~273}

   \author{
               M. Chernyakova \inst{1,} \inst{2} 
          \and A. Neronov \inst{1,} \inst{2} 
          \and	T. J.-L. Courvoisier \inst{1,} \inst{2}
           \and	M. T\"urler  \inst{1,} \inst{2}
           \and	S. Soldi   \inst{1,}  \inst{2}
           \and	V.~Beckmann   \inst{3}
           \and	P. Lubinski   \inst{4,} \inst{1} 
           \and	R.~Walter   \inst{1,}   \inst{2}
           \and	 K. L. Page   \inst{5}
           \and	 M. Stuhlinger   \inst{6}
            \and R. Staubert  \inst{7}
            \and I. M. McHardy  \inst{8}
             }

   \offprints{M.Chernyakova}

   \institute{
   INTEGRAL Science Data Center, Chemin d'\'Ecogia 16, 1290 Versoix, Switzerland
   \email {Masha.Chernyakova@obs.unige.ch}
   \and
    Geneva Observatory,  University of Geneva, 51 ch. des Maillettes, CH-1290 Sauverny, Switzerland
   \and
    NASA Goddard Space Flight Center,
Exploration of the Universe Division, Code 661,
Greenbelt, MD 20771, USA
   \and
   Nicolaus Copernicus Astronomical Center, Bartycka 18, 00-716 Warszawa, Poland 
   \and
  UK Swift Science Data Centre, 
Dept. of Physics \& Astronomy,   
University of Leicester,        
University Road,
Leicester LE1 7RH
    \and
     XMM-Newton Science Operations Centre,
European Space Astronomy Centre (ESAC) - Villafranca,
P.O. Box 50727 - 28080 Madrid - Spain 
   \and
    Institut f\"ur Astronomie und Astrophysik - Astronomie, 72076 T\"ubingen, Germany 
    \and
    School of Physics \& Astronomy
University of Southampton
Highfield, Southampton
SO17 1BJ, U.K.
   }

   \date{}

 
  \abstract
   {}
   {
    The aim of this paper is to study the evolution of the broadband
    spectrum of one of the brightest and nearest  quasars \thc.    
   }
   {
We analyze the data obtained during quasi-simultaneous \intgr\ and
\xmm\ monitoring of the blazar \thc\ in 2003--2005 in the UV, X-ray
and soft $\gamma$-ray bands and study the results in the context of 
the long-term evolution of the source.}
   {The 0.2-100~keV spectrum of the source is well fitted
by a combination of a soft cut-off power law and a hard power law.  No
improvement of the fit is achieved if one replaces the soft cut-off
power law by either a blackbody, or a disk reflection model. During the
observation period the source has reached the historically softest state in
the hard X-ray domain with a photon index $\Gamma=1.82\pm
0.01$. Comparing our data with available archived X-ray data from
previous years, we find a secular evolution of the
source toward softer X-ray emission (the photon index has increased by
$\Delta\Gamma\simeq 0.3-0.4$ over the last thirty years).  We argue that
existing theoretical models have to be significantly modified to account for the
observed spectral evolution of the source.}
   {}


   \maketitle

\section{Introduction}

3C 273 is a radio loud quasar, with a jet showing superluminal motion,
discovered at the very beginning of quasar research.  Being one of the
brightest and nearest (z=0.158) quasars, 3C~273 was intensively studied at 
different wavelengths (see \citet{courvoisier98} for a review).

In the X-ray band \thc\ is thought to be quite a unique source in the sense that
in its X-ray spectrum the ``blazar-like'' emission from the pc-scale jet mixes
with  the ``nuclear'' component, which reveals itself through the presence of 
a large blue-bump and a so-called  ``soft excess'' below $\sim
1$~keV. Both are typical for emission from the nuclei of Seyfert
galaxies. During the observation period discussed in this paper the
radio to millimeter  component, usually associated with a jet, has
been in its lowest state. Thus, our data provide an opportunity to 
study the ``nuclear'' component in more details and to test 
theoretical models for the emission of \thc.

Surprisingly, in spite of the low level of the ``jet-like'' contribution, our
data do not reveal the presence of typical ``Seyfert-like'' features in the
spectrum. In fact, the 0.2-100~keV spectrum is well fitted by a simple model
composed of two power laws or a combination of a hard power law with a soft
cut-off power law. We find that more sophisticated models, which include
a black-body for description of the soft excess, or a disk reflection component
give comparable or worse fits to the data, than the above simple model.
Moreover, the ``disentanglement'' of Seyfert and jet components of the X-ray
spectrum, as proposed by \citet{grandi04}, is not observed in our data set.\\

\citet{walter92} have found an anticorrelation between the X-ray spectral slope
and the ratio of X-ray to ultraviolet (UV) luminosity of \thc\ using quasi-simultaneous data
from EXOSAT, GINGA and IUE obtained in the 80-s. They have
interpreted this correlation as evidence that the X-ray emission is due to thermal 
Comptonization of the UV flux. The  presence of optical monitors
on-board of both \xmm\  and \intgr\ enables us to analyze correlations between the
big blue bump variability and the variability in X-rays.  Repeating the
analysis of \citet{walter92} with our data set we do not find the previously
observed X-ray slope -- UV/X-ray flux correlation.  The disappearance of this
correlation is even more surprising because of the appearance of a previously
unobserved correlation between the UV and the X-ray flux from the system. Taking
into account the fact that the typical variability time scales in the UV and
the X-ray bands are different (months in UV and days in X-ray), true simultaneity 
of the observations could be important. Our \xmm\ and \intgr\ observations are the
first observations in which UV and X-ray data were taken at the same time.

It is possible, however, that appearance/disappearance of the UV--X-ray
correlation reflects a real evolution of the spectral state of the source
during the last 30 years. Collecting the archival X-ray data for about 30 years
of observations, we find that the source indeed evolves towards a softer X-ray
state. We find that the source has reached its historically softest X-ray state
with photon index $\Gamma=1.82\pm 0.01$ in June 2003. This spectral slope is (by
$\Delta\Gamma \simeq 0.3-0.4$) softer than the typical slope observed in the 70-s and
80-s. This is definitely larger than the typical short time scale fluctuations of
the slope which are at a level of $\Delta\Gamma\simeq 0.1$.

It is important to note that spectral evolution of the source should not be
ignored in theoretical modelling of nuclear and jet emission from \thc. For
example, the very hard spectral state of the source in the late 70-s has led
\citet{lightman87} to the conclusion that ``compact source'' type models
dominated by pair production can not be applied to \thc\ in spite of its high
apparent compactness. However, this conclusion is not applicable for the
present state of the source. Similarly, the most popular model for the hard
X-ray spectrum of the source, inverse Compton emission from the pc-scale jet,
is satisfactory for the present state of the source, but cannot 
explain the $\Gamma\simeq 1.4$ spectral index observed in 70-s.

 The paper is organized as follows: in Section 2 we present the sequences of
observations and methods used for data reduction and analysis.  In section 3 we
present the results obtained, and discuss them in Section 4. We then give a
summary of our analysis in the last part of the paper.

\section[]{Observations and Data Analysis}

\subsection{\intgr\ observations}

Since the launch of \intgr\ \citep{winkler03} on  October 17, 2002, \thc\ was
regularly monitored. Details of the data used in the current analysis are
given in   Table \ref{intdata}. The preliminary analysis of the 2003 data
was done by \citet{courvoisier03}. A detailed study of the
multiwavelength spectrum of June 2004 was presented by \citet{turler06}.

\begin{table}
\caption{\intgr\ observations of \thc. \label{intdata}}
\begin{tabular}{@{}c@{\,}c@{\,}c@{\,}c@{\,}c@{}}
\\
\hline
Period& T$_{start}$& Rev&On Time &\ ISGRI/SPI/JEM-X\\
 & & & (ks)&Exposure(ks)\\
\hline
1a&2003-01-05&28&120.8&\\
1b&2003-01-11&30&10.6&\\
1c&2003-01-17&32&106.8&\\
1 &  &        & &182.5/182.4/12.4  \\ 
2a&2003-06-01&77&57.0&\\
2b&2003-06-04&78&52.6&\\ 
2c&2003-06-16&82&33.5&\\ 
2e&2003-07-06&89 -- 90&227.1 &\\
2f&2003-07-18&93&20.5 &\\
2g&2003-07-23&94&14.0 &\\
2 &  &  &     & 308.9/310.5/40.3\\  
3 &2004-01-01&148 -- 149&26.4&152.3/NA/13.8\\ 					                                   
4 &2004-06-23&207&  92.7& 67.1/NA/29.2\\					                                   
5a&2005-05-28&320 -- 321&278.3&\\ 
5b&2005-07-09&334&102.5&\\ 
5 &  &   &        &  269.1/367.2/68.7 \\
\hline
\end{tabular}
\label{integral_table}
\end{table}

During the \intgr\ observations \thc\ was all the time in a very low state,
with a flux of several milliCrabs (see Section \ref{spec}).  To obtain better
statistics for spectral analysis we have combined data taken within a few weeks,
which resulted in  five data sets. The combined exposure times
for each data set and each instrument on-board of \intgr\ are listed in Table
\ref{integral_table}.

The data for all \intgr\ instruments were analysed in a standard way with the
latest version of the off-line scientific analysis (OSA) package 
5.1 distributed by ISDC\footnote{ http://isdc.unige.ch/?Soft+download}
\citep{courvoisier03a}.

Most of the time, observations were done in the $5\times5$ dithering mode. The
staring mode was used only during revolutions 30, 148, and 149. It is therefore not 
possible to use SPI data during the third period. There are also
no SPI data during the fourth period, as the instrument was in the annealing
mode during this observation.

For the analysis of the \intgr/JEM-X observations we have selected science
windows for which the source was less than 3.5\deg\ from the center of the field of view. During the
first three periods JEM-X 2 was  operating, while during the fourth and
fifth period it was switched off, and data were taken with the JEM-X 1
telescope.

To minimize the influence of the flares in the surrounding background of the
charged particles we have not used pointings in which the detector countrate
exceeds the 
mean of the given period by more than 5\%.

\subsection{\xmm\ observations}

The log of the \xmm\ observations analyzed in this paper is presented in Table
\ref{xmmdata}. Among all the observations of the source done with  \xmm\ we
have selected those that were done while   \intgr\
was observing.

\begin{table}
\caption{\xmm\ observations of \thc. \label{xmmdata}}
\begin{tabular}{@{}c@{\,}c@{\,}c@{\,}c@{\,}c@{}}
\hline
Period&\ \ Date&\ \ Revolution&\ \ On Time &\ \ PN Exposure Time\\
 & & & (ks)&(ks)\\
\hline
1&2003-01-06&563&8.95&5.97\\
2&2003-07-07&655&58.56&40.60\\
3&2003-12-14&735&8.91&5.77\\
4&2004-06-30&835&62.91&13.93\\
5&2005-07-10&1023&28.11&19.33\\                   
\hline
\end{tabular}
\end{table}
\begin{figure*} \begin{center}
\includegraphics[width=\textwidth,angle=0]{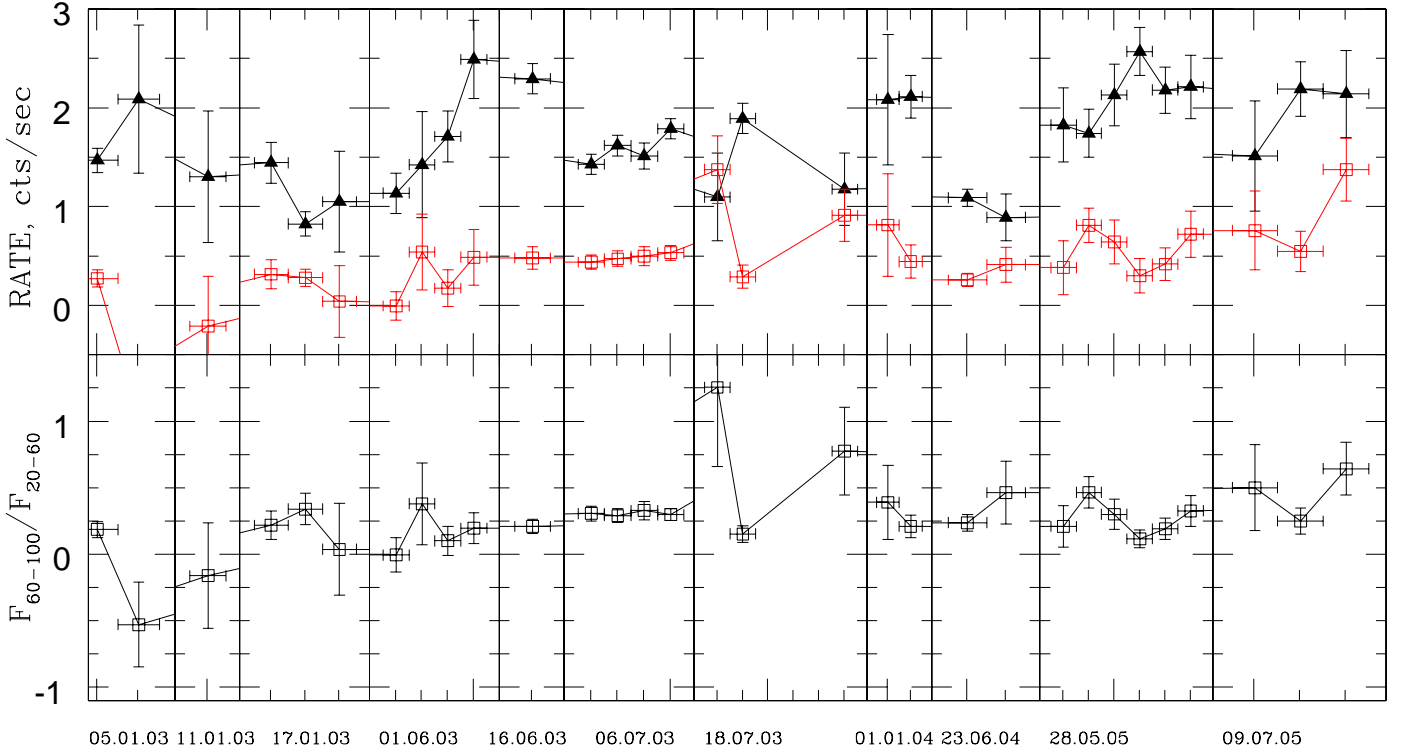}
\end{center} \caption{\textit{top panel:} 2003--2005 IBIS/ISGRI lightcurves of \thc in 
20-60 keV (upper curve, black) and 60-100 keV (lower curve, red). The data are
binned in 1-d intervals. For each time frame the date given at the bottom corresponds to the first data point,
tick interval is 1 day. \textit{bottom panel:} Hardness ratio 
F$_{(60-100\,keV)}$/F$_{(20-60\,keV)}$.} 
\label{lc_tot}
\end{figure*}

The \xmm\ Observation Data Files (ODFs) were obtained from the on-line Science
Archive\footnote{ http://xmm.vilspa.esa.es/external/xmm\_data\_acc/xsa/index.shtml}.
The data were processed and the event-lists filtered using {\sc xmmselect}
within the Science Analysis Software ({\sc sas}), version 6.5 and the most
up-to-date calibration files.   All EPIC observations were performed in the
Small Window Mode. In all observations the source was observed with the
MOS1, MOS2 and the PN detectors. As the PN detector, in contrast to MOS1 and 
MOS2, is not affected by pile-up in our observations we concentrate 
 our analysis on the PN data only.
Patterns 0-4 were used in the analysis.
Background spectra were extracted from areas of the sky close to \thc\
showing no obvious sources. ARFGEN and RMFGEN were then run to produce the 
corresponding ARF and RMF files.

Optical and ultraviolet fluxes were obtained with the optical monitor (OM) of
\emph{XMM-Newton} with \textit{omichain} following the instructions of the SAS
Watchout
page\footnote{ http://xmm.esac.esa.int/sas/documentation/watchout/uvflux.shtml}
and using the values provided for an AGN spectral type. The resulting magnitudes
are given in Table \ref{ommag}. No OM data for the selected filters are available for Period 3.

\begin{table}
\caption{Optical and UV  magnitudes of 3C 273 during each observational period.}
\label{ommag}																    
\begin{tabular}{@{}ccccc@{}}
\hline
&Period 1&Period 2&Period 4&Period 5\\
\hline
V&&12.53$\pm$0.01&12.69$\pm$0.01&12.57$\pm$0.01\\
B&&&12.95$\pm$0.01&12.86$\pm$0.01\\
U&&11.62$\pm$0.01&11.76$\pm$0.01&11.69$\pm$0.01\\
UVW1&11.32$\pm$0.01&11.28$\pm$0.01&11.44$\pm$0.01&11.30$\pm$0.01\\
UVM2&11.34$\pm$0.01&&11.37$\pm$0.01&11.19$\pm$0.01\\
UVW2&&11.14$\pm$0.01&11.36$\pm$0.01&11.16$\pm$0.01\\
\hline
\end{tabular}
\end{table}

\section[]{Results}


\subsection{Timing Analysis}

In the top panel of Figure \ref{lc_tot} we show the \thc\ IBIS/ISGRI lightcurves
in the 20-60 keV (black points) and in the 60-100 keV (red points) energy ranges. 
In this figure the data are binned in one day intervals. 
ISGRI data show that the hard X-ray emission from the system is variable on at least a
one day time scale. In order to investigate the nature
of this variability  we have produced the
``flux-flux'' diagram for the 20-60~keV and 60-100~keV energy bands, shown in the 
bottom right panel of Fig. \ref{fig:flfl}.  Large error bars of the single-day
flux measurements do not allow to see the general trend of the dependence between
the 20-60~keV and 60-100~keV flux. However, rebinning the data by building the
 average of three adjacent (along the  20-60~keV flux axis) original data points
 in the flux-flux plot we find that the whole data set has a roughly linear behaviour
(Spearman correlation coefficient is equal to $R = 0.89$, with a probability
of random coincidence $P_{rand} = 7 \times 10^{-3}$).
This  indicates that the variability of the source in the 20-100~keV band is mostly due to
changes in the normalization, but not in the shape of the spectrum.

\begin{figure} \begin{center}
\includegraphics[width=\columnwidth,angle=0]{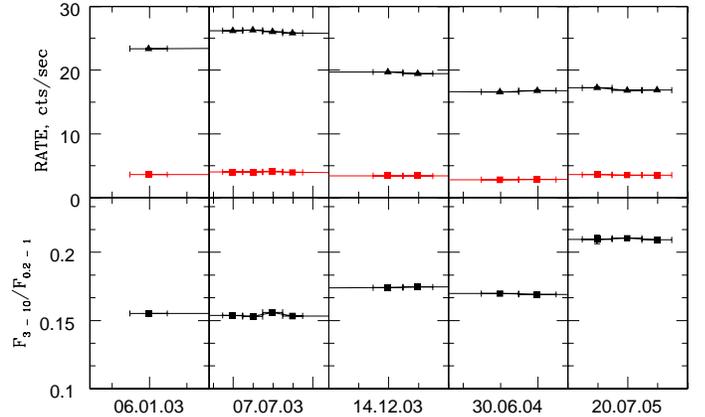} \end{center}
\caption{\textit{top panel:} 2003--2005 \textit{XMM}/PN lightcurves of \thc in 0.2--1\,keV
(upper curve, black) and 3--10\,keV (lower curve, red). The data are binned
in quarter of day intervals. \textit{bottom panel:} Hardness ratio  
F$_{(3-10\,keV)}$/F$_{(0.2-1\,keV)}$.}
\label{lc_xmm} 
\end{figure}

Fig. \ref{lc_xmm} shows the \xmm\ lightcurves in 0.2--1~keV (upper curve, black) and 3--10~keV
(lower curve, red) bands. One can see significant changes of the 0.2--1\,keV
flux from data set
to data set, but the variability at the time scale of single observation, $\sim
10$~ksec is negligible. The plot of the hardness ratio (bottom panel of Fig.
\ref{lc_xmm}) shows that the 0.2-10~keV spectrum hardens
toward the end of the observation period.

\subsection{Spectral Analysis \label{spec}}

\begin{figure}  \begin{center}
\includegraphics[width=\columnwidth]{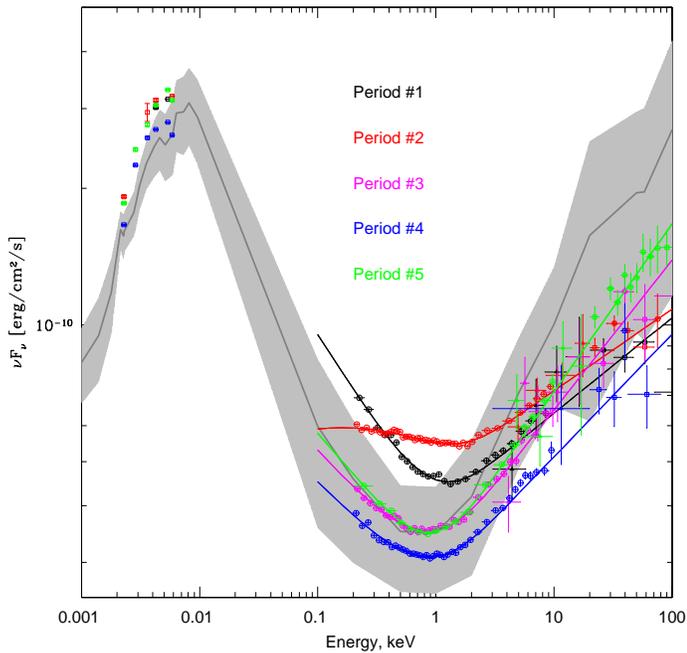}  
\end{center}
\caption{Broadband spectra of \thc\ for each period. Data are fitted with the 
simple phenomenological model described in Table \ref{pheno_param}.
The new data are compared to the average flux (grey line) and the 1-$\sigma$
  variability range (grey area) based on historic data of 3C\,273 \citep{turler99}.
} \label{allspec}
\end{figure}

Our \intgr\ and \xmm\ observations provide a very broad band (5 decades in
energy) set of quasi-simultaneous data which spans from the UV ($\sim 1$~eV) up to
the soft gamma-ray ($\sim 100$~keV) band. The  spectra of the source for all the
five selected data sets are shown in Fig. \ref{allspec}. Taking
into account a wide range of theoretical models proposed for the X-ray spectrum
of  \thc\ we adopt a ``simple to complex'' strategy in the spectral analysis 
and attempt first to find a ``minimal model'' which can provide  a satisfactory
representation of  the data and then consider how changing and/or adding 
various theoretically motivated components improves or worsens the fit. 
The hydrogen column density was fixed to the galactic value $N_H=1.79\times10^{20}cm^{-2}$ \citep{dickey90}.

\subsubsection{Phenomenological model}

Data in the 2--100 keV energy range are well fitted by a simple power law,
see Table \ref{1po_param}. In our analysis we have left the ISGRI, SPI and JEM-X 
intercalibration factors free to vary with respect to the PN camera. The
significant variation of the JEM-X intercalibration factor reflects a known
change of the gain parameters of the instrument. It is possible to fit all the 
data with the same SPI ($C_s=1.41)$ and ISGRI ($C_i=1.19$) intercalibration
factors, but we have preferred not to do this, as the \xmm\ and \intgr\ data 
are only quasi-simultaneous, and the different value of the flux measured by \xmm\ and \intgr\
can be due to the physical change of the source flux on a one month scale. The fit
of the second data set can be improved by the addition of a broad emission
line as discussed below. 

\begin{table}
  \begin{center}
    \caption{Parameters of the single power-law model in 2-100 keV
 energy range. The fit was done leaving free intercalibration factors for ISGRI
(C$_i$), SPI (C$_s$) and JEM-X (C$_j$) with respect to  the PN camera. The
parameter uncertainties correspond to 1 $\sigma$ level.}
\label{1po_param}
\begin{tabular}{@{}c@{\,}c@{\,}c@{\,}c@{\,}c@{\,}c@{\,}c@{}}
\hline
 &Period 1 &Period 2 &Period 3&Period 4&Period 5\\ 
\hline
\multicolumn{6}{c}{Power law}\\ 
$\Gamma$& $ 1.81^{+0.01}_{-0.01}$& $ 1.84^{+0.01}_{-0.01}$& $ 1.70^{+0.01}_{-0.01}$& $ 1.75^{+0.01}_{-0.01}$& $ 1.66^{+0.01}_{-0.01}$\\
$F_{2-10}^*$& $ 8.79^{+0.17}_{-0.16}$& $ 10.10^{+0.04}_{-0.07}$& $ 8.04^{+0.13}_{-0.12}$& $ 6.72^{+0.10}_{-0.09}$& $ 8.54^{+0.09}_{-0.09}$\\
\multicolumn{6}{c}{Intercalibration factors}\\ 
$C_{j}$&$ 1.11^{+0.09}_{-0.09}$ &$ 1.35^{+0.05}_{-0.05}$ &$ 1.33^{+0.09}_{-0.09}$ &$ 0.30^{+0.11}_{-0.11}$ &$ 0.67^{+0.05}_{-0.05}$ \\
$C_{i}$&$ 0.92^{+0.05}_{-0.05}$ &$ 1.22^{+0.03}_{-0.03}$ &$ 1.42^{+0.11}_{-0.11}$ &$ 0.98^{+0.08}_{-0.07}$ &$ 1.23^{+0.04}_{-0.04}$ \\
$C_{s}$&$ 1.39^{+0.26}_{-0.25}$ &$ 1.49^{+0.16}_{-0.16}$ &--&--&$ 1.67^{+0.15}_{-0.14}$ \\
\multicolumn{6}{c}{Statistics}\\ 
$\chi^2$&920 &1673 &950 &1139 &1419 \\
(dof)&(920) &(1632) &(1153) &(1205) &(1390) \\
\hline 
\end{tabular}
\end{center}
\begin{flushleft} 
$^\star$ In units of $10^{-11}$ erg cm$^{-2}$ s$^{-1}$. 
 
\end{flushleft} 
\end{table} 
 

\begin{figure} \begin{center}
\includegraphics[width=\columnwidth,angle=0]{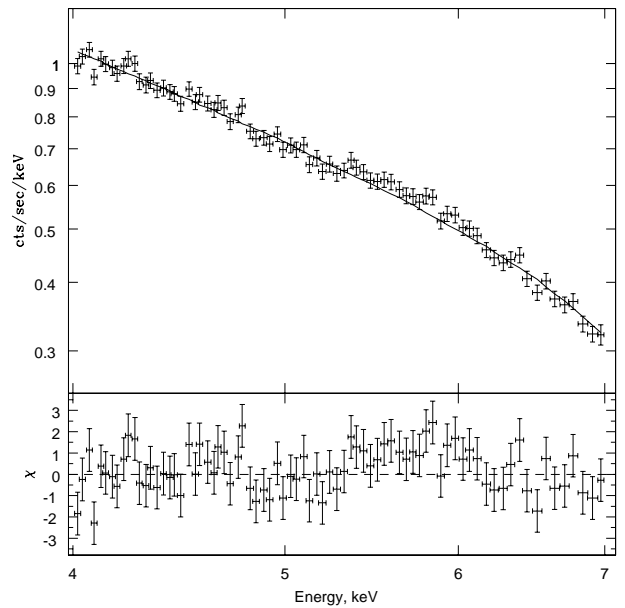} \end{center}
\caption{{A \thc\ spectrum in the 4-7 keV energy band as observed with \textit{XMM}/PN in July 2003 (second data set).
Data are fitted with a single power law. A broad iron line is marginally visible in the residuals (see text).}}
\label{feline} 
\end{figure}

The simplest satisfactory fit to the 0.2-100~keV data can be achieved with a
model consisting of two power laws. A soft power law with a photon index $\Gamma_1$
dominates in the energy band below 2~keV and describes the so-called "soft
excess". A hard power law with photon index $\Gamma_2$  
dominates in the hard X-ray band. An F-test shows that a slightly 
better fit to the data can be achieved if one uses a cut-off power law instead
of the single power law for the description of the soft emission.

Surprisingly, we find that already this simple model provides a very good
 fit to the combined \intgr\ and \xmm\ data in all the observations except the
second one. In the second data set the addition of a broad redshifted emission line at
$E_{Fe}=6.65\pm0.07$~keV in the rest frame of the object makes the fit acceptable. This line can be 
identified with  a fluorescent K$_\alpha$  iron line. A broad iron line was seen in previous observations of \thc\ by
different instruments, see e.g. \citet{yaqoob00}. The large width and marginal
detection of the line {(see Figure \ref{feline})} do not allow us to firmly distinguish  ionised and
non-ionised iron.  
In all other spectra the addition of an iron line does not significantly improve
the fit.
For completeness we show the degradation of the  fit if no line
is included. The complete list of the fit parameters for our
phenomenological model is given in Table \ref{pheno_param}. 


\begin{table}
  \begin{center}
    \caption{Parameters of the phenomenological model fit to the data.
 When the uncertainties are not quoted,  the value of the parameter was  fixed.}
\label{pheno_param}
\begin{tabular}{@{}c@{\,}c@{\,}c@{\,}c@{\,}c@{\,}c@{}}
\hline
 &Period 1 &Period 2 &Period 3&Period 4&Period 5\\
\hline
\multicolumn{6}{c}{Cut-off power law describing Soft excess}\\
$\Gamma_1$& $2.46^{+0.14}_{-0.16}$ &$ 2.02^{+0.08}_{-0.08}$&2.37$^{+0.12}_{-0.13}$&$2.36^{+0.12}_{-0.13}$&$2.32^{+0.10}_{-0.11}$\\
$E_1$&      $0.65^{+0.20}_{-0.14}$ &$ 0.52^{+0.06}_{-0.05}$&0.73$^{+0.24}_{-0.15}$&$0.65^{+0.16}_{-0.12}$&$0.58^{+0.10}_{-0.08}$\\
$F^\star_{0.2-1}$&3.60$^{+1.08}_{-0.76}$&$2.11^{+0.16}_{-0.26} $&2.34$^{+0.59}_{-0.41}$&1.89$^{+0.49}_{-0.35}$&$2.36^{+0.24}_{-0.39}$\\
\multicolumn{6}{c}{Second power law describing hard tail}\\
$\Gamma_2$& 1.79$^{+0.02}_{-0.02}$ &$ 1.82^{+0.01}_{-0.01}$&$1.67^{+0.02}_{-0.01}$&$1.73^{+0.01}_{-0.01}$&$1.63^{+0.01}_{-0.01}$\\
$F_{2-10}^\star$&8.76$^{+0.19}_{-0.21}$&$10.03^{+0.06}_{-0.10}$&8.02$ ^{+0.22}_{-0.17}$&$6.70^{+0.10}_{-0.12}$&$8.53^{+0.08}_{-0.10}$\\
\multicolumn{6}{c}{Gaussian line}\\
E$_{Fe}     $&6.65$                $ &6.65$^{+0.07}_{-0.07}$&6.65$                $&6.65                   &6.65$                  $\\
$\sigma_{Fe}$&1.98$^{+0.55}_{-0.39}$ &0.31$^{+0.07}_{-0.06}$&0.39$^{+0.22}_{-0.12}$&0.50$^{+0.92}_{-0.50}$&2.35$ ^{+2.97}_{-1.30}$\\
I$_{Fe}^{\star\star}$&42.50$^{+32.32}_{-12.80}$&6.1$^{+1.3}_{-1.4}$&5.29$^{+3.31}_{-2.51}$&$3.14^{+3.91}_{-2.79}$&17.14$^{+5.83}_{-12.31}$\\
\multicolumn{6}{c}{Intercalibration factors}\\
C$_j$&1.11$^{+0.09}_{-0.09}$&$ 1.33^{+0.05}_{-0.05}$&1.31$^{+0.09}_{-0.09}$&0.37$ ^{+0.10}_{-0.10}$&0.66$ ^{+0.05}_{-0.05}$\\
C$_i$&0.87$^{+0.05}_{-0.05}$&$ 1.16^{+0.03}_{-0.03}$&1.38$^{+0.11}_{-0.10}$&0.97$ ^{+0.07}_{-0.07}$&1.19$ ^{+0.37}_{-0.35}$\\
C$_s$&1.32$^{+0.24}_{-0.24}$&$ 1.43^{+0.15}_{-0.15}$&          --          &    --                 &1.62$ ^{+0.14}_{-0.13}$\\
\multicolumn{6}{c}{Statistics}\\
$\chi^2$&   1305 & 2174 & 1360 & 1594 & 1832\\
(df)&      (1308)&(2022)&(1519)&(1634)&(1778)\\
\multicolumn{6}{c}{Statistics without iron line taken into account}\\
$\chi^2_{nl}$&1316 & 2204 & 1364 & 1596 & 1836\\
(df)&      (1310)&(2025)&(1521)&(1636)&(1780)\\
\hline
\end{tabular}
\end{center}
\begin{flushleft}
$^\star$ in $10^{-11}$ erg cm$^{-2}$ s$^{-1}$ units.

$^{\star\star\star}$  in $10^{-5}$ photons cm$^{-2}$ s$^{-1}$  units.
\end{flushleft}
\end{table}

\subsubsection{Modelling of the soft excess}

The ``soft excess'' below 1~keV is typical of Seyfert galaxies. Several
competing models for the physical origin of this component are
available. In particular, the most simple ``phenomenological'' models of the
soft excess include a (multi-temperature) black body component. However,
the origin of such blackbody component is not clear, because the inferred
temperature is usually too high to be accounted for standard accretion disks. More
complicated models include e.g. reflection of external X-ray emission from the
disk. Attempts to fit the soft excess in our \xmm\ data with different models
give the following results. 

\begin{table} \begin{center} \caption{Parameters of the model which includes a
black body to describe the soft excess.} \label{bb_param}
\begin{tabular}{@{}c@{\,}c@{\,}c@{\,}c@{\,}c@{\,}c@{}} 
\hline &Period 1 &Period2 &Period 3&Period 4&Period 5\\  
\hline \multicolumn{6}{c}{Black body}\\  
T(eV)&$78.9^{+1.5}_{-1.6}$&$ 87.2^{+0.8}_{-0.7}$&$ 79.0^{+1.8}_{-1.2}$&$78.0^{+2.0}_{-1.2}$&$ 82.0^{+1.0}_{-1.0}$\\ 
$F_{0.2-1}^*$& $ 1.93^{+0.05}_{-0.06}$& $1.08^{+0.01}_{-0.01}$& $ 1.21^{+0.03}_{-0.03}$& $ 0.94^{+0.02}_{-0.03}$& $1.42^{+0.02}_{-0.03}$\\
 \multicolumn{6}{c}{Power law}\\ 
$\Gamma$& $1.892^{+0.007}_{-0.006}$& $ 1.877^{+0.001}_{-0.001}$& $1.787^{+0.005}_{-0.006}$& $ 1.815^{+0.005}_{-0.006}$& $1.719^{+0.004}_{-0.003}$\\ 
$F_{2-10}^*$& $ 8.48^{+0.05}_{-0.04}$& $
9.88^{+0.01}_{-0.01}$& $ 7.75^{+0.03}_{-0.04}$& $ 6.52^{+0.02}_{-0.03}$& $
8.32^{+0.03}_{-0.02}$\\ 
\multicolumn{6}{c}{Gaussian line}\\  $E_{Fe}$& $ 6.65$&
$  6.72^{+0.04}_{-0.06}$& $ 6.65$& $ 6.65$& $ 6.65$\\ $\sigma_{Fe}$& $ 0.10$& $
0.10$& $ 0.10$& $ 0.10$& $ 0.10$\\ $n_{Fe}^{**}$& $ 4.56^{+1.64}_{-1.57}$& $
4.89^{+0.37}_{-1.10}$& $ 3.54^{+1.29}_{-1.26}$& $ 2.49^{+0.94}_{-0.95}$& $
2.35^{+0.90}_{-0.85}$\\ 
\multicolumn{6}{c}{Intercalibration factors}\\ 
$C_{j}$&$ 1.18^{+0.10}_{-0.10}$ &$ 1.39^{+0.05}_{-0.05}$ &$
1.41^{+0.10}_{-0.10}$ &$ 0.40^{+0.12}_{-0.11}$ &$ 0.70^{+0.06}_{-0.06}$ \\
$C_{i}$&$ 1.13^{+0.06}_{-0.05}$ &$ 1.34^{+0.03}_{-0.03}$ &$
1.80^{+0.12}_{-0.13}$ &$ 1.21^{+0.08}_{-0.08}$ &$ 1.46^{+0.04}_{-0.03}$ \\
$C_{s}$&$ 1.72^{+0.31}_{-0.31}$ &$ 1.65^{+0.18}_{-0.18}$ &--&--&$
1.98^{+0.17}_{-0.17}$ \\
 \multicolumn{6}{c}{Statistics}\\  $\chi^2$&1445 &2659
&1543 &1799 &2161 \\ (dof)&(1310) &(2024) &(1515) &(1630) &(1780) \\
\multicolumn{6}{c}{Statistics without iron line taken into account}\\  $\chi^2_{nl}$&1453 &2710 &1551
&1807 &2169 \\ (dof)&(1311) &(2026) &(1516) &(1631) &(1781) \\ \hline 
\end{tabular} 
\end{center} 
\begin{flushleft}  
$^\star$ In units of $10^{-11}$erg cm$^{-2}$ s$^{-1}$. 

$^{\star\star}$ In units of $10^{-5}$ photons keV$^{-1}$ cm$^{-2}$ s$^{-1}$. 
\end{flushleft} 
\end{table} 

{A black body model} for the soft excess does not provide an acceptable
fit to the data. Best fit parameters for this model are listed in Table
\ref{bb_param}.The disk reflection component calculated using the {\tt kdblur} and {\tt reflion} 
models of XSPEC \citep{ross05,crummy06} provides a better description
of the soft excess, than the blackbody model. Among the parameters of the 
{\tt reflion} model we fit only the  spectral slope of the power law emission
of the central source, fluxes of the direct and reflected components, and the
ionization parameter of the gas $\xi$ ($\xi=4\pi F/n_H$, where $F$ is the 
illuminating energy flux, and $n_H$ is the hydrogen number density in the 
illuminated layer; in the  {\tt reflion} model $\xi$ is restricted to
$1<\xi<10000$ erg cm s$^{-1}$). Parameters of the model fit are given in 
Table \ref{refl_param}. In this Table we also report the 
measured ratio  $R$ of the reflected flux to the total flux expressed in percents.  

 \begin{table}
  \begin{center}
    \caption{{Parameters of the  model consisting of emission from the central 
    source reflected from the relativistically-blurred photoionized disc and a jet} }
\label{refl_param}
\begin{tabular}{@{}c@{\,}c@{\,}c@{\,}c@{\,}c@{\,}c@{}}
\hline
 &Period 1 &Period 2 &Period 3&Period 4&Period 5\\ 
\hline
\multicolumn{6}{c}{{Direct and reflected emission from the central source}}\\ 
$\xi^+$&$ 2977^{+1565}_{-1126}$&$ 10000^{+0}_{-248}$&$ 3170^{+3360}_{-1433}$&$ 10000^{+0}_{-2453}$&$ 2570^{+2391}_{-410}$\\
$F_{0.2-1}^*$& $ 5.78^{+0.71}_{-0.72}$& $ 4.31^{+0.31}_{-0.36}$& $ 4.59^{+0.60}_{-0.59}$& $ 4.05^{+0.63}_{-0.39}$& $ 3.88^{+0.16}_{-0.35}$\\
$\Gamma_1$& $ 2.83^{+0.07}_{-0.08}$& $ 2.56^{+0.01}_{-0.02}$& $ 2.56^{+0.08}_{-0.08}$& $ 2.48^{+0.07}_{-0.09}$& $ 2.76^{+0.04}_{-0.01}$\\
$R$(\%)& $ 2.39^{+3.98}_{-1.00}$& $ 20.02^{+179.51}_{-2.39}$& $ 2.61^{+2.01}_{-1.56}$& $ 11.38^{+10.69}_{-7.11}$& $ 2.69^{+3.41}_{-1.15}$\\
\multicolumn{6}{c}{{Power law describing Jet emission}}\\ 
$\Gamma_2$& $ 1.70^{+0.03}_{-0.02}$& $ 1.75^{+0.01}_{-0.02}$& $ 1.53^{+0.04}_{-0.03}$& $ 1.55^{+0.05}_{-0.06}$& $ 1.54^{+0.03}_{-0.02}$\\
$F_{2-10}^{*}$& $ 8.05^{+0.27}_{-0.57}$& $ 9.23^{+0.25}_{-0.36}$& $ 6.88^{+0.60}_{-0.75}$& $ 5.76^{+2.63}_{-2.80}$& $ 8.02^{+2.68}_{-0.38}$\\
\multicolumn{6}{c}{Gaussian line}\\ 
$E_{Fe}$& $ 6.65$& $ 6.73^{+0.05}_{-0.06}$& $ 6.65$& $ 6.65$& $ 6.65$\\
$n_{Fe}^{**}$& $ 3.81^{+1.57}_{-1.56}$& $ 3.06^{+0.62}_{-0.63}$& $ 2.28^{+1.37}_{-1.35}$& $ 0.20^{+0.93}_{-0.20}$& $ 1.42^{+0.94}_{-0.86}$\\
\multicolumn{6}{c}{Intercalibration factors}\\ 
$C_{j}$&$ 1.10^{+0.04}_{-0.05}$ &$ 1.19^{+0.02}_{-0.02}$ &$ 1.28^{+0.05}_{-0.54}$ &$ 0.35^{+0.05}_{-0.05}$ &$ 0.66^{+0.04}_{-0.02}$ \\
$C_{i}$&$ 0.77^{+0.09}_{-0.09}$ &$ 1.17^{+0.04}_{-0.05}$ &$ 1.12^{+0.09}_{-0.09}$ &$ 0.76^{+0.10}_{-0.10}$ &$ 1.08^{+0.05}_{-0.05}$ \\
$C_{s}$&$ 1.16^{+0.21}_{-0.21}$ &$ 1.30^{+0.14}_{-0.14}$ &--&--&$ 1.35^{+0.12}_{-0.11}$ \\
\multicolumn{6}{c}{Statistics}\\ 
$\chi^2$&1399 &2239 &1349 &1516 &1819 \\
(dof)&(1410) &(2002) &(1507) &(1542) &(1758) \\
\multicolumn{6}{c}{Statistics without iron line taken into account}\\ 
$\chi^2_{nl}$&1405 &2262 &1353 &1516 &1822 \\
(dof)&(1412) &(2004) &(1509) &(1543) &(1759) \\
\hline 
\end{tabular}
\end{center}
\begin{flushleft}
$^+$ in units of erg cm s$^{-1}$.
 
$^\star$ in units of $10^{-11}$ erg cm$^{-2}$ s$^{-1}$. 

$^{\star\star}$ in units of $10^{-4}$ photons keV$^{-1}$ cm$^{-2}$ s$^{-1}$. 
\end{flushleft} 
\end{table} 


One should note, however, that the amount of reflection found in each data set,
except for data set number 2, is at the level of several percent. Thus, a better
fit (compared to blackbody) to the data is explained not by the careful account
of reflection features, but by a large contribution of the direct power law
illuminating the disk, so that the disk reflection model is not significantly
different from the previous phenomenological model. To summarize, we find that  
neither the black body nor the disk reflection provide better descriptions of
the soft excess than the (cut-off) power law.

Compton reflection is characterized by a "reflection hump" at about
30~keV. To look for the presence of a reflection component in the spectrum
we have also tried to fit the spectra above 2~keV with either a simple
power law or a direct plus reflected power law (the {\tt reflion} model of XSPEC)
and compare the  quality of the fits. Although we find that the addition of
the reflected component at the level of about 20\% improves the quality of the
fits, the F-test shows that the probability that this improvement is 
coincidental is at the level of 10\%, so that no definitive conclusion about
the presence of a reflection hump in the spectrum can be made. One should note
that the largest uncertainty is due to the fact that the ISGRI instrument is
sensitive only above $\sim 20$~keV, while \xmm\ data are available only below
$\sim 10$~keV and, in fact, there is a degeneracy between 
the normalization of the reflection hump and the ISGRI -- \xmm\ intercalibration
factor.

A combination of black-body and reflection components, suggested as a
best-fit model for the Beppo-SAX data by \citet{grandi04} does not change the
above conclusion {(see below)}, leaving the (cut-off) power law to be the best model for the
soft excess.

\begin{figure*}
\begin{center}
\includegraphics[width=\textwidth]{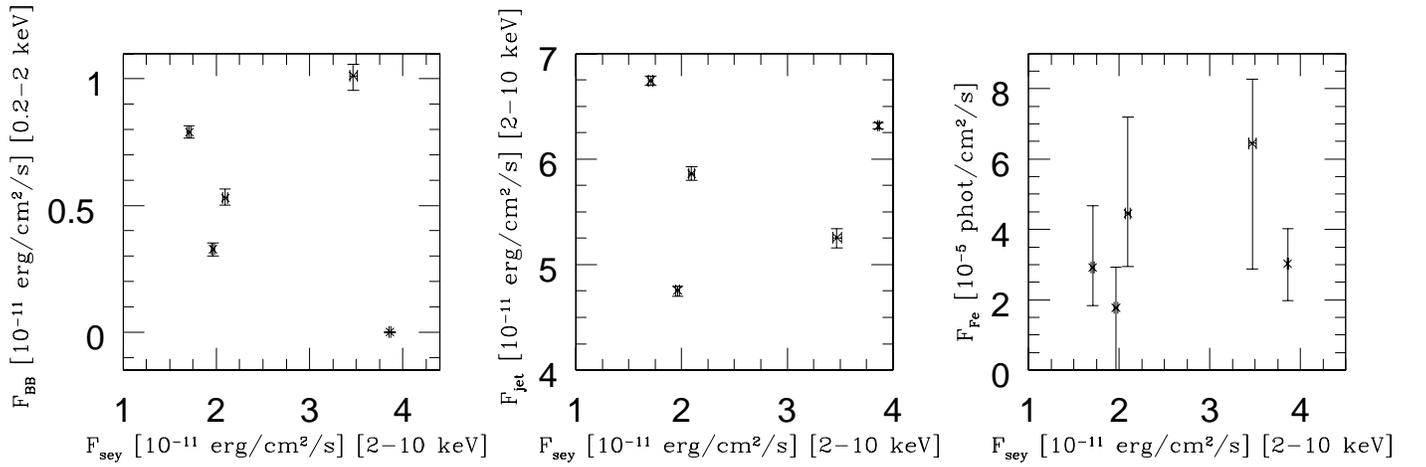}
\end{center}
\caption{The dependance of the 0.2-2 keV black-body emission (\textit{left}),
the 2-10 keV jet like flux (\textit{middle}), and the iron line flux (\textit{right})
on the  2-10 keV Seyfert-like flux. Data points are taken from Table \ref{pexrav_param}    }
\label{npex_cor}
\end{figure*}

\subsubsection{Disentangling the Jet and Seyfert contributions}
\citet{grandi04} (GP) have put forward a strong claim that Seyfert and ``jet''
components of the X-ray spectrum could be disentangled in the case of \thc\ so
that the observed variations of the X-ray spectral shape are successfully
described by  variations in the normalizations of the template ``jet'' and
``Seyfert'' components. We have  tested this hypothesis with our
data. For this we repeat the analysis of \citet{grandi04}. First, we find
the ``template'' models for Seyfert and jet components by fitting
simultaneously all the 5 data sets with the same model which consists of a hard
power law (jet) and black-body plus partially reflected cut-off power law (Seyfert
component). The best fit is achieved with the following parameters:
a black body temperature $T_{bb}=75$~eV, a Seyfert component power law slope
$\Gamma_s=2.25$, a reflection component fraction $RC=0.77$, a jet power law slope
$\Gamma_j=1.5$, an inclination of the disk $\phi=18^\circ$. We assumed  solar
abundances of elements and {fixed the position and  width of the iron line at 
the values found in the second data set: $E_{Fe}=6.65$, $\sigma_{Fe}=0.3$~keV}. Then we froze the above
parameters and left only the normalizations of the black body, reflection and jet
power law components variable. With such a ``restricted model'' we fit each data
set separately and find the normalizations of the three ``template''
components. The fluxes in each component found in this way are listed in Table
\ref{pexrav_param}.
\begin{table}
  \begin{center}
    \caption{Black body, power law, Seyfert-like component 
and Gaussian line fit to PN, JEM-X, SPI and ISGRI 3C 273 spectra.
The fit was done leaving free intercalibration factors
  for ISGRI (C$_i$), SPI (C$_s$) and JEM-X (C$_j$) with respect to 
  the PN camera. The parameter uncertainties correspond to 1 $\sigma$ level.
   }
\label{pexrav_param}
\begin{tabular}{@{}c@{\,}ccccc@{}}
\hline
 &Period 1 &Period 2 &Period 3&Period 4&Period 5\\ 
\hline
\multicolumn{6}{c}{Black body}\\ 
$F_1^*$& $ 1.01^{+0.05}_{-0.06}$& $ <0.004$& $ 0.53^{+0.03}_{-0.03}$& $ 0.33^{+0.02}_{-0.03}$& $ 0.79^{+0.03}_{-0.02}$\\
\multicolumn{6}{c}{Power law}\\ 
$F_2^*$& $ 5.25^{+0.09}_{-0.09}$& $ 6.32^{+0.03}_{-0.03}$& $ 5.86^{+0.07}_{-0.06}$& $ 4.76^{+0.04}_{-0.05}$& $ 6.74^{+0.05}_{-0.04}$\\
\multicolumn{6}{c}{Seyfert component}\\ 
$F_s^{*}$& $ 3.47^{+0.04}_{-0.04}$& $ 3.86^{+0.01}_{-0.01}$& $ 2.10^{+0.03}_{-0.03}$& $ 1.96^{+0.02}_{-0.02}$& $ 1.71^{+0.02}_{-0.02}$\\
\multicolumn{6}{c}{Gaussian line}\\ 
$n_{Fe}^{**}$& $ 6.45^{+1.82}_{-3.58}$& $ 3.03^{+0.99}_{-1.06}$& $ 4.45^{+2.74}_{-1.51}$& $ 1.78^{+1.15}_{-1.78}$& $ 2.92^{+1.76}_{-1.08}$\\
\multicolumn{6}{c}{Intercalibration factors}\\ 
$C_{j}$&$ 1.09^{+0.09}_{-0.09}$ &$ 1.26^{+0.05}_{-0.05}$ &$ 1.39^{+0.10}_{-0.10}$ &$ 0.28^{+0.11}_{-0.11}$ &$ 0.65^{+0.05}_{-0.05}$ \\
$C_{i}$&$ 0.68^{+0.03}_{-0.03}$ &$ 0.80^{+0.02}_{-0.02}$ &$ 1.16^{+0.09}_{-0.09}$ &$ 0.74^{+0.05}_{-0.05}$ &$ 1.05^{+0.02}_{-0.02}$ \\
$C_{s}$&$ 1.01^{+0.18}_{-0.19}$ &$ 0.97^{+0.11}_{-0.11}$ &--&--&$ 1.42^{+0.12}_{-0.12}$ \\
\multicolumn{6}{c}{Statistics}\\ 
$\chi^2$&1340 &2342 &1383 &1561 &1913 \\
(dof)&(1311) &(2026) &(1531) &(1565) &(1781) \\
\multicolumn{6}{c}{Statistics without iron line taken into account}\\ 
$\chi^2_{nl}$&1345 &2350 &1389 &1562 &1918 \\
(dof)&(1312) &(2027) &(1532) &(1566) &(1782) \\
\hline 
\end{tabular}
\end{center}
\begin{flushleft} 
$^\star$ in units of $10^{-11}$ erg cm$^{-2}$ s$^{-1}$. 

$^{\star\star}$ in units of $10^{-5}$ photons keV$^{-1}$ cm$^{-2}$ s$^{-1}$. 
\end{flushleft} 
\end{table} 


After this analysis, and following \citet{grandi04},  we study the
correlations between e.g. Seyfert component
and iron line strength or between fluxes in Seyfert and black body component,
which were found in the  Beppo-SAX data.  Fig. \ref{npex_cor} (which is an
analog of Fig. 1 of GP) shows that no such correlations are
present in our data set. We discuss the implications of this result below.\\

\subsubsection{{Multi band flux and spectral slope correlations}} 
\begin{figure}
\begin{center}
\includegraphics[width=\columnwidth]{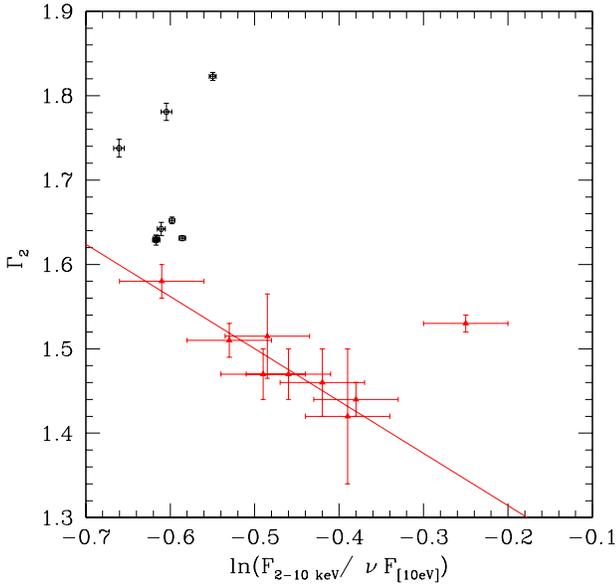}
\end{center}
\caption{{2 -- 10 keV spectral photon index of the phemenological model } as a function of the logarithm 
of the ratio of the 2--10\,keV to ultraviolet fluxes. Old \textit{Ginga} and \textit{EXOSAT}
data are shown with red triangles. The line indicates the best correlation found in \citet{walter92}. New \xmm\ data are shown with black circles. We have extrapolated the UV spectrum 
obtained with the XMM optical monitor up to 10 eV using the time-averaged UV
spectrum of \thc.} 
\label{comptcor}
\end{figure}

Comptonization of big blue bump photons was proposed as a plausible
mechanism of production of the hard X-ray emission from \thc\ by
\citet{walter92}. The motivation for this model was a linear correlation between
the X-ray spectral slope above 2 keV and the logarithm of the ratio of the 
2--10\,keV flux to the ultraviolet and soft excess count rates found in \textit{EXOSAT},
\textit{GINGA} and \textit{IUE} data. The presence of the UV camera on board of \xmm\ and of the
Optical Monitor Camera (OMC) on board of \textit{INTEGRAL} enables us to test  
this correlation between the  UV and X-ray fluxes and slopes. Fig. \ref{comptcor} shows the values
of the power law index in 2--10\,keV band, $\Gamma_2$, plotted against the
logarithm of the ratio of X-ray and UV fluxes (F$_{2-10\,keV}$/$\nu$F$_{10
eV}$). \footnote{We have extrapolated the UV spectrum obtained with the \xmm\ optical
monitor up to 10 eV using the time-averaged UV spectrum of \thc.}

Figure \ref{comptcor} clearly shows that the correlations seen in EXOSAT/GINGA X-ray data
and IUE UV data are not present in our data set. For completeness we have
extended the analysis for the whole set of publicly available \xmm\ data on
\thc\ to produce this Figure (for details see Soldi et al., in preparation).

\begin{figure} 
\begin{center}
\includegraphics[width=\columnwidth]{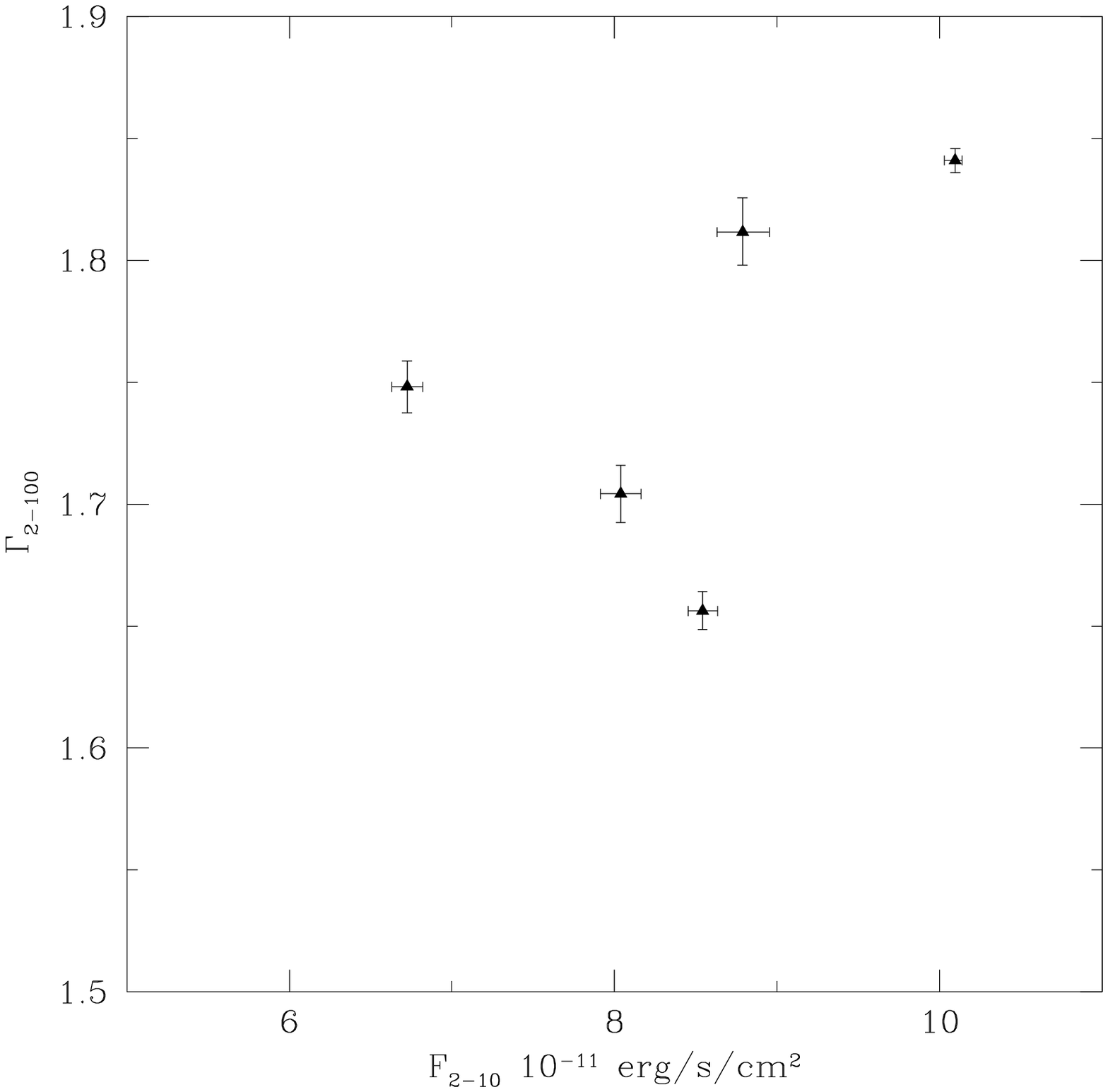}
\end{center}
\caption{The dependence of the hard X-ray photon index {(derived from fitting the data in  2-100
keV energy range with a single power law, see table \ref{1po_param})} on the 2--10\,keV X-ray flux.}
\label{marc}
\end{figure}
\begin{figure*}
\begin{center}
\includegraphics[width=0.75\textwidth,angle=0]{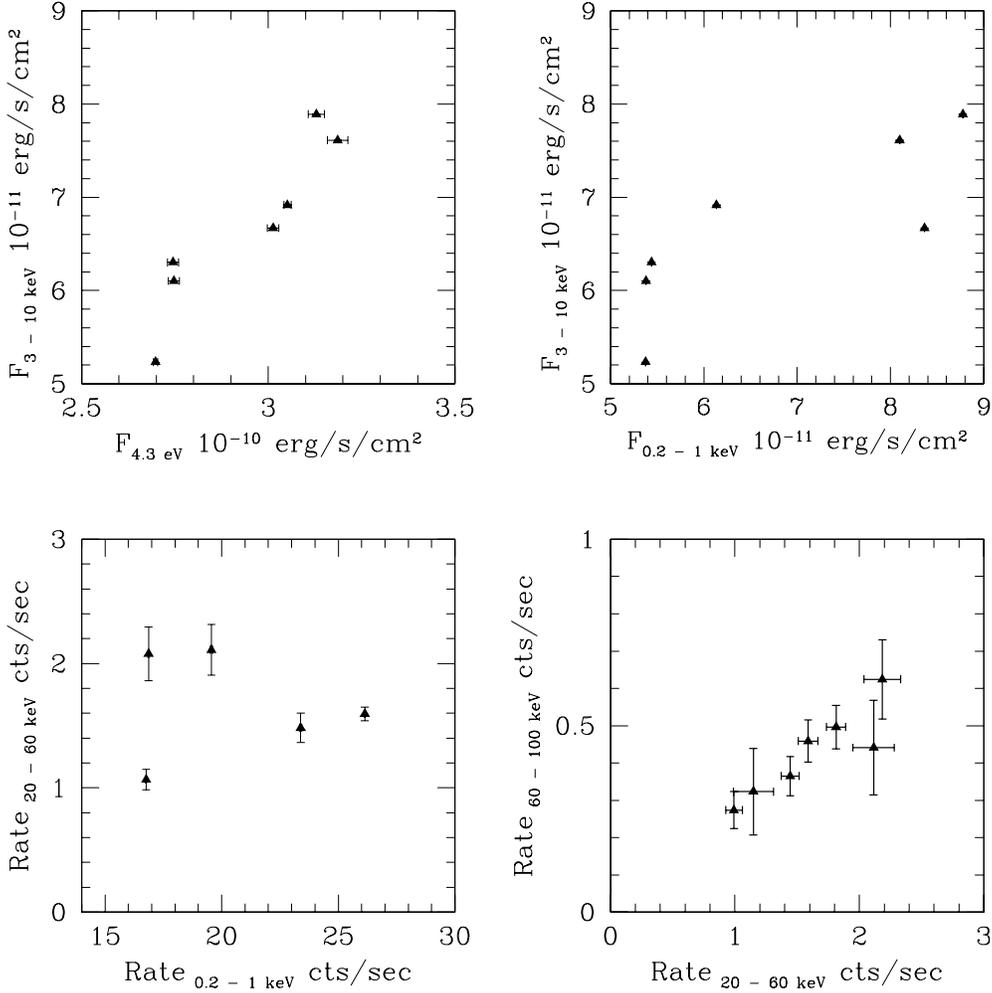}
\end{center}
\caption{Flux-flux plots for different energy bands. On the right bottom panel 
one day averaged \intgr\ data are shown. To reduce
the scatter the data were rebinned by  building the
 average of three adjacent original data  points in x direction.}
\label{fig:flfl}
\end{figure*}

In  Fig. \ref{comptcor} two important facts are apparent. First, the scatter of
the \xmm\ data points along the x-axis is much smaller than that of the EXOSAT/GINGA
and IUE points.  This means 
that in the recent past the X-ray and UV fluxes have varied in a correlated
way. Left top panel of Fig.  \ref{fig:flfl} shows that this is indeed the case 
(Spearman correlation coefficient is equal to $R = 0.93$, with a probability
of random coincidence $P_{rand} = 3 \times 10^{-3}$).
This is a quite surprising result, because previous searches for such correlations
gave negative results. However another correlation  typical for Seyfert
galaxies  between the hard X-ray spectral index
$\Gamma_{2}$ and the X-ray flux (\citet{zdziarski03}, and
references therein), is not observed (see Fig.~\ref{marc}).  A similar result was found by 
\citet{page04}.

Second, the values of the hard X-ray spectral index in \xmm\ observations are
systematically higher than those of the EXOSAT/GINGA data.  This indicates that
the spectral state of the source has evolved over the 30 years of X-ray
observations (in the next section we discuss this in details).  This may be linked to the
``disappearance'' of the correlations seen in EXOSAT, GINGA and IUE data and
``appearance'' of X-ray -- UV flux correlations not detected before.
	  
{Having found a previously undetected correlation between 
UV and X-ray flux, we have searched for the presence of additional correlations, such as between the
0.2-1~keV and 3-10~keV fluxes (top right panel of Fig. \ref{fig:flfl}), 0.2-1~keV and 20-60~keV fluxes 
(bottom left panel of Fig. \ref{fig:flfl}) and 20-60~keV and 60-100~keV fluxes (bottom right panel of Fig. 
\ref{fig:flfl}). One can see a clear correlation between 20-60~keV and 60-100~keV fluxes, indicating that both
bands are dominated by one and the same physical component (e.g. powerlaw emission from the jet). On the other side, no or only 
marginal 
correlation in the two other cases shows that several physical mechanisms contribute to the emission in these energy bands. } 

\section{Long-term evolution of the  X-ray spectrum of \thc}

To test the evolution of the X-ray spectrum over the last 30 year
period we have collected the archival X-ray data from different
missions. The evolution of the hard (2-10~keV) spectral index found
from the archival 
data\footnote{Available from 3C 273's Database: http://isdc.unige.ch/3c273/}
\citep{turler99} is shown in Fig. \ref{gamhist}.  One
clearly sees both significant short-term variations of the spectral
index with a scatter of $\Delta\Gamma\simeq 0.1$
and a  long-term trend of softening of the spectrum by as much as
$\Delta\Gamma\simeq 0.4$ over the last 30~yr. It is interesting to
note that during our 2003-2005 monitoring campaign the source has
reached its historically softest hard X-ray spectrum characterized by
a photon index $\Gamma=1.82\pm 0.01$. It is possible that still
softer spectral states  were found at the very beginning of X-ray
observations (UHURU, ARIEL5/C and OSO8/A data points), but the uncertanity of
those measurements are too large to be conclusive. 

\begin{figure*}
\begin{center}
\includegraphics[width=\textwidth]{6285fig9.epsi}
\end{center}
\caption{Evolution of the flux density at 1 keV, as extrapolated from the fit done above 2 keV, in $10^{-2}$photons/keV/cm$^2/s$ (top panel) and hard X-ray spectral index 
(bottom panel) over the 30 years
  of observations. \textit{UHURU}, \textit{Ariel 5}/C, \textit{OSO 8}/A, \textit{HEAO-1}, \textit{HEAO-2}/MPC, and \textit{BALL}/AIT 
   data points are taken from \citet{malaguti94};
  \textit{EXOSAT} and \textit{Ginga} data are taken from \citet{turner90}; \textit{ASCA} data are from \citet{cappi98}; 
  \textit{RXTE}, \textit{Beppo}-SAX, and part of \textit{XMM-Newton} 
   data are from  Soldi et al. (in preparation).   }
\label{gamhist}
\end{figure*}

\section{Discussion}

Most of the theoretical models of \thc\ suggest that the source spectrum
contains  two major contributions, one from a pc-scale jet and the other from a
Seyfert-like nucleus. The very low millimeter flux measured during our observations suggested that the jet contribution has
 reached its historical minimum (see \citet{turler06}). However, this did not result in the clear appearance of the Seyfert-like
contribution. {This means either that the  Seyfert-like contribution has decreased simultaneously with the jet contribution (which, because of the 
difference in the size scales and production sites can happen only by a chance coincidence), or that the observed emission just can not be decomposed 
into  Seyfert and jet contributions.}

Our data show that
the separation of Seyfert and jet contributions to the
spectrum could not be done by fixing a "template" for each of them.  
Namely, although the typical ranges of variations of soft and hard X-ray fluxes
in our data set are similar to the ones of {\it Beppo-SAX} data set, the "template" models
for the Seyfert, jet and black body components found in our data set are significantly different. 
In particular, the additional black body component
is  not required by the data and thus the resulting black body flux shown in the left panel of Fig. \ref{npex_cor} is
 systematically lower than the values reported in \citet{grandi04}. Our data show that the correlation between the black body and 
Seyfert components is absent at such low black body flux values. 
The same is true for the non-observation of the correlation between the Seyfert component flux
and the strength of the iron line: since the addition of the iron line is not statistically required in any data sets except for the second 
one, 
no correlation is seen at such low  iron line flux values. However, it is important to note 
that no direct comparison between our Fig. \ref{npex_cor} and Fig. 1 of \citet{grandi04} can be done, because of the differences in the 
template models derived from  Beppo-SAX  and \intgr\ -\xmm\ data sets.

In fact, the analysis of the secular evolution of the source 
presented above shows that both Seyfert and jet spectra evolve with time.
This can be illustrated by the results presented in Fig. \ref{comptcor}, where 
we plot the hard X-ray spectral slope as a 
function of  the ratio of the X-ray power law flux to the UV flux. 
This  ratio is apart from a constant factor the ratio of
the Comptonized flux to the
source photon flux in a model in which the X-ray power law is due to the
Comptonization of the soft photons by a thermal electron population.  The
points given by the red crosses with large error bars 
were obtained by the EXOSAT and IUE missions in
the 80-s and were analysed by \citet{walter92} who deduced that in
the frame of the thermal Comptonization model, the electron temperature was
about  1MeV or somewhat less. They also deduced that the optical depth of the
Comptonizing medium was 0.1--0.2 and that it covered a couple of percent of the
soft photon source. 

The \xmm\ data  shown in Figure \ref{comptcor} by the black points 
with small error bars clearly do not match the correlation
found in earlier years. This is particularly true for the spectra obtained during periods 1 and 2,
when the soft X-ray emission was much brighter than at the other epochs, while the ultraviolet
flux was similar. This indicates a much harder blue bump for these epochs. More photons were 
therefore available to Compton cool the hot electrons than inferred from the ultraviolet flux shown in Figure 5.
This therefore leads to the softening in the X-ray emission. Alternatively, one can also infer a modification
of the Comptonization region when the source is very weak. The recent data also show that the hard X-ray flux 
is correlated with the soft flux (see Figure 8), an effect not observed at earlier epochs \citep{courvoisier90}.


The correlation between  the optical/UV and the X-ray flux is characteristic
for Seyfert galaxies.  This relationship is particularly clear in the long-term
behaviour of NGC~5548 \citep{uttley03}, but was also seen in other Seyfert
galaxies like for instance NGC~4051 (\citet{shemmer03}, and references
therein). The data of \thc\ presented here do also show the appearance of such
a trend  (see Fig.~\ref{fig:flfl}), apparently suggesting a Seyfert-like
behavior of 3C~273 in 2003--2005. However, it is clear from the above
discussion that in  the case of \thc\ conclusions of this sort should be taken
with caution: there is no clear way to disentangle the Seyfert and jet
contributions. For example,  another characteristic of Seyfert
galaxies, a correlation between the hard X-ray spectral index
$\Gamma_{2}$ and the X-ray flux (\citet{zdziarski03}, and
references therein), is not observed (see Fig.~\ref{marc}). 

The secular evolution of the hard X-ray spectral index over the last 30 years, 
clearly seen in Fig. \ref{gamhist}, shows that the jet contribution also evolves
with time.  Whether the hard X-ray flux is generated in the jet, or 
it is due to the nuclear related physics (i.e. from the "compact
source"{ \citep{lightman87}}) is still completely undecided.
"Jet" and "compact source" models 
differ in their approach to the so-called compactness problem.
The essence of the compactness problem is that the  naively estimated optical depth
of the source for $\gamma$-rays with energies $E_\gamma$, such that pairs are produced in 
the interactions with soft photons of energies $\epsilon_s\ge
2m_e^2c^4/E_\gamma$, is higher than 1. The
"compact source" models self-consistently account for the effect of the pair
production cascade in the source {\citep{lightman87}}. The "jet" models get rid of the compactness
problem by assuming relativistic beaming of the $\gamma$-ray emission.

The observed evolution of the hard X-ray spectral index from $\Gamma\simeq 1.4$
to $\Gamma\simeq 1.8$ over the last 30 years could help to distinguish between
the two possible models of the high-energy emission. The point is that although
the present-day values $\Gamma>1.5$ are  easily explained in both models,  the
very hard spectrum observed is the late 70-s is not. In both
models the gamma-rays are produced via inverse Compton emission from
high-energy electrons. The hard spectral index $\Gamma\simeq 1.4$ observed in
70-s and 80-s (see Fig. \ref{gamhist}) is too hard for the optically thin 
inverse Compton emission from an electron distribution cooled via synchrotron and
inverse Compton energy loss which can produce only spectra with $\Gamma\ge 1.5$.

This problem can, however, be relaxed if one assumes that 
in addition to 
electrons the emission region contains also protons. If the main source of the 
seed photons for the inverse Compton scattering is UV photons from 
the big blue bump, the hard X-ray
inverse Compton emission originates from a population of electrons with moderate
energies $E_e\sim 1-10$~MeV. In this energy range cooling could be dominated by
ionization, rather than inverse Compton/synchrotron losses. The Coulomb loss
leads to the hardening of the electron spectrum, compared to the injection
spectrum by $\Delta\Gamma_e\simeq -1$. This will result in a hardening of the 
inverse Compton
spectrum  below the so-called "Coulomb break" (see e.g.
\citet{aharonian04}). A simple estimate shows that if the 
proton density in the gamma-ray emission region is
 $n_p\sim 10^{10}\mbox{ cm}^{-3}$, the Coulomb break is situated in the 
 hard X-ray band which can provide the explanation of the harder than 
 $\Gamma= 1.5$ spectrum of the source in late 70-s. High matter density in the
 emission region obviously favors the "compact source" models compared to the
 jet models: the assumption about a large
($10^{10}$~cm$^{-3}$) matter densities at the pc distances from the nucleus
needed to explain the hard spectrum in the jet model framework looks somewhat
extreme (for
comparison, the density of dust and molecular gas in the Central Molecular Zone
 (CMZ) of the Galaxy is estimated to be $10^4$~cm$^{-3}$, see e.g.  \citet{morris96}).
The conclusion that the observed evolution of the hard X-ray spectral index
favors the compact source models should also be taken with caution. The reason
for this is that the presence of a Seyfert-like contribution to
the hard X-ray spectrum can affect the value of the spectral index.

\begin{figure}
\includegraphics[width=1.1\columnwidth]{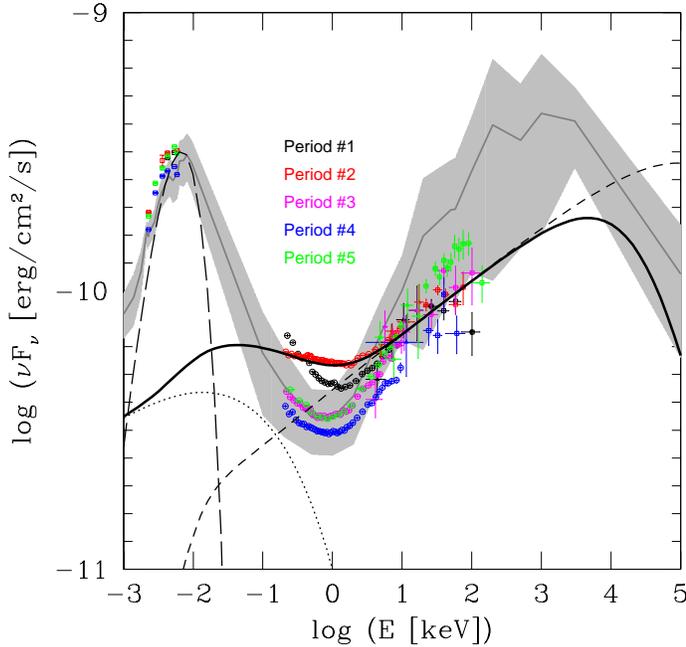}
\caption{Synchrotron-external-Compton fits to the data for the observation
period 2 in the "compact source"
type model. Solid line: the total spectrum, short dashed line: the inverse
Compton spectrum without the account of the pair production. Dotted
line: synchrotron spectrum. Long dashed line: the seed
photon spectrum for the inverse Compton emission. Similar to Fig. \ref{allspec}
the  average flux (grey line) and the 1-$\sigma$
  variability range (grey area) is shown.}
\label{fig:spectrum}
\end{figure}

If the gamma-ray emission of \thc\ originates in a compact
source the synchrotron radiation from electrons
which upscatter the blue bump photons to the GeV energy band could contribute
also to the soft X-ray spectrum of the source.  
The magnetic field strength in the $\gamma$-ray emission region is poorly
constrained by the observations (see, however, \citet{courvoisier88} for restrictions 
on the magnetic field strength during flaring activity of the source in the late
80-s). Assuming that the energy density of magnetic field in the compact emission 
region is comparable to the energy density of radiation, 
\begin{equation}
U_{ph}\simeq 2\times 10^{2}
\left[\frac{L}{10^{46}\mbox{erg/s}}\right]\left[\frac{
10^{16}\mbox{cm}}{R}\right]^2\mbox{erg/cm}^{3}
\end{equation}
(we estimate the size of emission region to be $R\sim 10^{16}$~cm from the
observed intra-day variability of the source) one obtains an estimate
\begin{equation}
B\simeq 10^2\left[\frac{L}{10^{46}\mbox{erg/s}}\right]^{1/2}\left[\frac{
10^{16}\mbox{cm}}{R}\right]\mbox{ G}
\end{equation}
The synchrotron radiation from 1-10~GeV electrons in such magnetic field 
is emitted at the energies
\begin{equation}
\epsilon_s=3\left[\frac{B}{10^{2}\mbox{ G}}\right]\left[\frac{E_e}{10\mbox{ GeV}}\right]^2\mbox{ keV}
\end{equation}
The non-observation of a cut-off in the GeV band {\citep{collmar00}} implies that the
high-energy cut-off in the electron spectrum is at least at energies
$E_e\ge 1$~GeV which, in turn, implies that the cut-off in the synchrotron
spectrum should be at least in the soft X-ray band.
Taking into account that the best fit to the "soft X-ray excess" in our data is
given by a simple cut-off power law model, we find that it is possible that
  the observed soft
X-ray excess in the spectrum of \thc\ is the high-energy tail of the synchrotron
radiation from the compact core of the source.  As an example, we show in Fig.
\ref{fig:spectrum} a synchrotron-inverse Compton fit to the broad band 
spectrum of the source for the observation period 2 (with the historically 
steepest spectrum in the hard X-ray band). The parameters used for the model fit
are: a magnetic field $B=33$~G, a size of emission region $R=10^{16}$~cm, 
a spectral index and cut-off in the electron
spectrum are $\Gamma_e=2.6$ and $E_{cut}=10$~GeV, respectively. The gamma-ray 
spectrum has a break at $E=10$~MeV due to pair production. For comparison,
short-dashed line in this figure shows the spectrum of inverse Compton emission
without taking into account  the pair production.

\section{Summary}

In this paper we have presented the results of monitoring of the
quasar \thc\ over the period of 2003--2005 over the broad energy
band, spanning from optical/UV up to soft $\gamma$-rays with \xmm\ and
\intgr.  Our monitoring  campaign gave some surprising results. In particular, 
in our data we find little support for the previously observed correlations 
between certain characteristics of the source, such as the ones found in Beppo-SAX data by 
\citet{grandi04} (Fig. \ref{npex_cor}), and in  GINGA/EXOSAT data by
\citet{walter92}(Fig. \ref{comptcor}).  Moreover, we note the appearance of a new correlation between the 
UV and the X-ray flux from \thc, which was not present before. This 
points to the fact that the source undergoes a secular evolution. The secular evolution of the 
source is evident from the  study of the evolution of the X-ray spectral index during the last 30 
years of observations (Fig. \ref{gamhist}).  
  We argue that
existing theoretical models have to be significantly modified to account for the
observed spectral evolution of the source.

\section{Acknowledgments}
Authors are grateful to the unknown referee for helpful comments.
PL acknowledges the support of KBN grant 1P03D01128.

 \bibliographystyle{aa}  
\bibliography{biblio} 

\begin{thebibliography}{25}
\expandafter\ifx\csname natexlab\endcsname\relax\def\natexlab#1{#1}\fi

\bibitem[{{Aharonian}(2004)}]{aharonian04}
{Aharonian}, F.~A. 2004, {Very high energy cosmic gamma radiation: a crucial
  window on the extreme Universe} (published by World Scientific Publishing)

\bibitem[{{Cappi} {et~al.}(1998){Cappi}, {Matsuoka}, {Otani}, \&
  {Leighly}}]{cappi98}
{Cappi}, M., {Matsuoka}, M., {Otani}, C., \& {Leighly}, K.~M. 1998, \pasj, 50,
  213

\bibitem[{{Collmar} {et~al.}(2000){Collmar}, {Reimer}, {Bennett}, {Bloemen},
  {Hermsen}, {Lichti}, {Ryan}, {Sch{\"o}nfelder}, {Steinle}, {Williams}, \&
  {B{\"o}ttcher}}]{collmar00}
{Collmar}, W., {Reimer}, O., {Bennett}, K., {et~al.} 2000, \aap, 354, 513

\bibitem[{{Courvoisier}(1998)}]{courvoisier98}
{Courvoisier}, T.~J.-L. 1998, \aapr, 9, 1

\bibitem[{{Courvoisier} {et~al.}(2003{\natexlab{a}}){Courvoisier}, {Beckmann},
  {Bourban}, {Chenevez}, {Chernyakova}, {Deluit}, {Favre}, {Grindlay}, {Lund},
  {O'Brien}, {Page}, {Produit}, {T{\"u}rler}, {Turner}, {Staubert},
  {Stuhlinger}, {Walter}, \& {Zdziarski}}]{courvoisier03}
{Courvoisier}, T.~J.-L., {Beckmann}, V., {Bourban}, G., {et~al.}
  2003{\natexlab{a}}, \aap, 411, L343

\bibitem[{{Courvoisier} {et~al.}(1990){Courvoisier}, {Robson}, {Blecha},
  {Bouchet}, {Falomo}, {Maisack}, {Staubert}, {Terasranta}, {Turner},
  {Valtaoja}, {Walter}, \& {Wamsteker}}]{courvoisier90}
{Courvoisier}, T.~J.~L., {Robson}, E.~I., {Blecha}, A., {et~al.} 1990, \aap,
  234, 73

\bibitem[{{Courvoisier} {et~al.}(1988){Courvoisier}, {Robson}, {Hughes},
  {Blecha}, {Bouchet}, {Krisciunas}, \& {Schwarz}}]{courvoisier88}
{Courvoisier}, T.~J.-L., {Robson}, E.~I., {Hughes}, D.~H., {et~al.} 1988, \nat,
  335, 330

\bibitem[{{Courvoisier} {et~al.}(2003{\natexlab{b}}){Courvoisier}, {Walter},
  {Beckmann}, {Dean}, {Dubath}, {Hudec}, {Kretschmar}, {Mereghetti},
  {Montmerle}, {Mowlavi}, {Paltani}, {Preite Martinez}, {Produit}, {Staubert},
  {Strong}, {Swings}, {Westergaard}, {White}, {Winkler}, \&
  {Zdziarski}}]{courvoisier03a}
{Courvoisier}, T.~J.-L., {Walter}, R., {Beckmann}, V., {et~al.}
  2003{\natexlab{b}}, \aap, 411, L53

\bibitem[{{Crummy} {et~al.}(2006){Crummy}, {Fabian}, {Gallo}, \&
  {Ross}}]{crummy06}
{Crummy}, J., {Fabian}, A.~C., {Gallo}, L., \& {Ross}, R.~R. 2006, \mnras, 365,
  1067

\bibitem[{{Dickey} \& {Lockman}(1990)}]{dickey90}
{Dickey}, J.~M. \& {Lockman}, F.~J. 1990, \araa, 28, 215

\bibitem[{{Grandi} \& {Palumbo}(2004)}]{grandi04}
{Grandi}, P. \& {Palumbo}, G.~G.~C. 2004, Science, 306, 998

\bibitem[{{Lightman} \& {Zdziarski}(1987)}]{lightman87}
{Lightman}, A.~P. \& {Zdziarski}, A.~A. 1987, \apj, 319, 643

\bibitem[{{Malaguti} {et~al.}(1994){Malaguti}, {Bassani}, \&
  {Caroli}}]{malaguti94}
{Malaguti}, G., {Bassani}, L., \& {Caroli}, E. 1994, \apjs, 94, 517

\bibitem[{{Morris} \& {Serabyn}(1996)}]{morris96}
{Morris}, M. \& {Serabyn}, E. 1996, \araa, 34, 645

\bibitem[{{Page} {et~al.}(2004){Page}, {Turner}, {Done}, {O'Brien}, {Reeves},
  {Sembay}, \& {Stuhlinger}}]{page04}
{Page}, K.~L., {Turner}, M.~J.~L., {Done}, C., {et~al.} 2004, \mnras, 349, 57

\bibitem[{{Ross} \& {Fabian}(2005)}]{ross05}
{Ross}, R.~R. \& {Fabian}, A.~C. 2005, \mnras, 358, 211

\bibitem[{{Shemmer} {et~al.}(2003){Shemmer}, {Uttley}, {Netzer}, \&
  {McHardy}}]{shemmer03}
{Shemmer}, O., {Uttley}, P., {Netzer}, H., \& {McHardy}, I.~M. 2003, \mnras,
  343, 1341

\bibitem[{{T{\"u}rler} {et~al.}(2006){T{\"u}rler}, {Chernyakova},
  {Courvoisier}, {Foellmi}, {Aller}, {Aller}, {Kraus}, {Krichbaum},
  {L{\"a}hteenm{\"a}ki}, {Marscher}, {McHardy}, {O'Brien}, {Page}, {Popescu},
  {Robson}, {Tornikoski}, \& {Ungerechts}}]{turler06}
{T{\"u}rler}, M., {Chernyakova}, M., {Courvoisier}, T.~J.-L., {et~al.} 2006,
  \aap, 451, L1

\bibitem[{{T{\"u}rler} {et~al.}(1999){T{\"u}rler}, {Paltani}, {Courvoisier},
  {Aller}, {Aller}, {Blecha}, {Bouchet}, {Lainela}, {McHardy}, {Robson},
  {Stevens}, {Ter{\"a}sranta}, {Tornikoski}, {Ulrich}, {Waltman}, {Wamsteker},
  \& {Wright}}]{turler99}
{T{\"u}rler}, M., {Paltani}, S., {Courvoisier}, T.~J.-L., {et~al.} 1999, \aaps,
  134, 89

\bibitem[{{Turner} {et~al.}(1990){Turner}, {Williams}, {Courvoisier},
  {Stewart}, {Nandra}, {Pounds}, {Ohashi}, {Makishima}, {Inoue}, {Kii},
  {Makino}, {Hayashida}, {Tanaka}, {Takano}, \& {Koyama}}]{turner90}
{Turner}, M.~J.~L., {Williams}, O.~R., {Courvoisier}, T.~J.~L., {et~al.} 1990,
  \mnras, 244, 310

\bibitem[{{Uttley} {et~al.}(2003){Uttley}, {Edelson}, {McHardy}, {Peterson}, \&
  {Markowitz}}]{uttley03}
{Uttley}, P., {Edelson}, R., {McHardy}, I.~M., {Peterson}, B.~M., \&
  {Markowitz}, A. 2003, \apjl, 584, L53

\bibitem[{{Walter} \& {Courvoisier}(1992)}]{walter92}
{Walter}, R. \& {Courvoisier}, T.~J.-L. 1992, \aap, 258, 255

\bibitem[{{Winkler} {et~al.}(2003){Winkler}, {Courvoisier}, {Di Cocco},
  {Gehrels}, {Gim{\'e}nez}, {Grebenev}, {Hermsen}, {Mas-Hesse}, {Lebrun},
  {Lund}, {Palumbo}, {Paul}, {Roques}, {Schnopper}, {Sch{\"o}nfelder},
  {Sunyaev}, {Teegarden}, {Ubertini}, {Vedrenne}, \& {Dean}}]{winkler03}
{Winkler}, C., {Courvoisier}, T.~J.-L., {Di Cocco}, G., {et~al.} 2003, \aap,
  411, L1

\bibitem[{{Yaqoob} \& {Serlemitsos}(2000)}]{yaqoob00}
{Yaqoob}, T. \& {Serlemitsos}, P. 2000, \apjl, 544, L95

\bibitem[{{Zdziarski} {et~al.}(2003){Zdziarski}, {Lubi{\'n}ski}, {Gilfanov}, \&
  {Revnivtsev}}]{zdziarski03}
{Zdziarski}, A.~A., {Lubi{\'n}ski}, P., {Gilfanov}, M., \& {Revnivtsev}, M.
  2003, \mnras, 342, 355

\end{thebibliography}

\label{lastpage}

\end{document}